\documentclass[aps,prd,10pt,twocolumn,amsmath,floats,floatfix,nofootinbib,superscriptaddress,preprintnumbers]{revtex4-2}

\usepackage{soul}
\usepackage{silence} 
\usepackage{hyperref}
\usepackage{graphicx}
\usepackage{orcidlink} 
\usepackage[makeroom]{cancel}
\usepackage{enumitem}
\usepackage{amssymb,amsmath,mathrsfs,bm,bbm,mathtools} 
\usepackage{amsthm}
\definecolor{ao(english)}{rgb}{0.0, 0.5, 0.0}
\hypersetup{colorlinks=true,linkcolor=blue,urlcolor=blue,citecolor=ao(english)}
\usepackage{stackrel}
\usepackage{multirow}
\usepackage[normalem]{ulem} 

\newtheorem*{theorem*}{Theorem}

\global\arraycolsep=1truept
\def\cob{\color{blue}}

\renewcommand{\leq}{\leqslant}
\renewcommand{\geq}{\geqslant}
\newcommand{\Eq}[1]{(\ref{#1})}
\newcommand{\Eqq}[1]{Eq.~(\ref{#1})}

\newcommand{\be}{\begin{equation}}
\newcommand{\ee}{\end{equation}}
\newcommand{\ben}{\begin{equation*}}
\newcommand{\een}{\end{equation*}}
\newcommand{\ba}{\begin{eqnarray}}
\newcommand{\ea}{\end{eqnarray}}
\newcommand{\ban}{\begin{eqnarray*}}
\newcommand{\ean}{\end{eqnarray*}}
\newcommand{\bs}{\begin{subequations}}
\newcommand{\es}{\end{subequations}}
\newcommand{\bc}{\begin{center}}
\newcommand{\ec}{\end{center}}
\newcommand\nn{\nonumber\\}

\def\cA{\mathcal{A}}
\def\cB{\mathcal{B}}
\def\cC{\mathcal{C}}

\def\cL{\mathcal{L}}

\def\cN{\mathcal{N}}
\def\cO{\mathcal{O}}

\def\cS{\mathcal{S}}

\def\p{\partial}
\def\Re{\text{Re}}

\def\a{\alpha}

\def\de{\delta}

\def\g{\gamma}
\def\la{\lambda}

\def\e{\epsilon}
\def\ve{\varepsilon}
\def\Om{\Omega}
\def\om{\omega}
\def\G{\Gamma}
\def\t{\tau}

\def\vp{\varphi}

\def\B{\Box}

\def\rme{{\rm e}}
\def\rmd{{\rm d}}
\def\rmi{{\rm i}}

\newcommand{\au}[2]{#1.~#2}
\newcommand{\arX}[1]{\href{http://arxiv.org/abs/#1}{{\cob arXiv:#1}}}
\newcommand{\oarX}[1]{\href{http://arxiv.org/abs/#1}{{\cob #1}}}

\newcommand{\book}[5]{\emph{#1} (#2, #3, #4, #5)}

\newcommand{\doin}[6]{\href{http://dx.doi.org/#1}{{\cob #2 #3 {\bf #4}, #5 (#6)}}}
\newcommand{\doinn}[5]{\href{http://dx.doi.org/#1}{{\cob #2 {\bf #3}, #4 (#5)}}}
\newcommand{\doij}[5]{\href{http://dx.doi.org/#1}{{\cob #2 #3 (#5) #4}}}

\newcommand{\ndoinn}[5]{\href{#1}{{\cob #2 {\bf #3}, #4 (#5)}}}

\newcommand{\procm}[6]{in \emph{#1}, edited by #2 (#3, #4, #5, #6)}
\newcommand{\procsinm}[5]{in \emph{#1}, edited by #2 (#3, #4, #5)}
\newcommand{\tia}[1]{{#1},}

\def\lst{\ell_*}

\begin{document}

\title{Diffusion method in field theories with fakeons} 

\author{Gianluca Calcagni\,\orcidlink{0000-0003-2631-4588}}
\email{g.calcagni@csic.es}
\affiliation{Instituto de Estructura de la Materia, CSIC,\\ Serrano 121, Madrid 28006, Spain}

\begin{abstract}
We adapt the diffusion method employed in fundamentally nonlocal field theories to determine the number of initial conditions for the classicized dynamics of unitary field theories with fakeons, characterized by inverse powers of the d'Alembertian operator $\Box$. We show that this number is two and we recover all the results obtained with a direct calculation elsewhere, including explicit solutions of linear toy models. Possible applications to nonlocal gravity are discussed.
\end{abstract}

\preprint{\doin{10.1103/hk2t-4qlz}{PHYSICAL REVIEW}{D}{113}{025006}{2026} [\arX{2510.15157}]}

\maketitle


\section{Introduction}\label{intro}

Nonlocality in quantum field theory (QFT) is the presence of operators (called form factors) characterized by infinitely many derivatives or, which is equivalent, represented by convolutions over all the points of spacetime (see, e.g., \cite{BasiBeneito:2022wux}). It can appear as a byproduct both of the loop expansion in standard perturbative local field theory and effective field theory \cite{Coleman:1973jx,Barvinsky:1985an,Barvinsky:1987uw,Manohar:1996cq,Avramidi:2000bm,Shapiro:2008sf,Donoghue:2014yha} and of the classicization procedure in theories with purely virtual particles or ``fakeons'' \cite{Anselmi:2017yux,Anselmi:2017lia,Anselmi:2018bra,Anselmi:2019rxg,Anselmi:2020tqo,Anselmi:2021hab,Anselmi:2022toe,Anselmi:2025uzj}. Nonlocality is also a fundamental feature of some beyond-standard settings with intrinsic scales, such as infrared nonlocal gravity \cite{Barvinsky:2003kg,Deser:2007jk,Belgacem:2017cqo,Belgacem:2020pdz}, nonlocal quantum gravity (NLQG) \cite{BasiBeneito:2022wux,Modesto:2017sdr,Buoninfante:2022ild,Koshelev:2023elc}, fractional field theories and fractional quantum gravity (FQG) \cite{Trinchero:2012zn,Trinchero:2017weu,Trinchero:2018gwe,Calcagni:2021ipd,Calcagni:2021ljs,Calcagni:2021aap,Calcagni:2022shb,Calcagni:2025wnn}. In some of its forms, nonlocality has shown to have the potential to improve renormalizability of QFTs (NLQG and FQG), to preserve unitarity in otherwise unstable quantum models (NLQG) and to give rise to unique phenomenology both at large and at small scales (infrared nonlocal gravity and NLQG).

Perhaps the major hurdle of nonlocal systems is the notorious difficulty in solving their dynamics consistently. While linear nonlocal equations can usually be solved with field redefinitions and direct methods \cite{Pais:1950za,Barnaby:2007ve}, nonlinear nonlocal systems do not admit such tricks and are thus more difficult to deal with. Moreover, the representation and handling of nonlocal operators can be so delicate that, if not done properly, can lead to all sorts of inconsistencies \cite{Eliezer:1989cr,LH00,Moeller:2002vx,LH08}. With the purpose of solving these problems, a method based on one or more diffusionlike equations was proposed \cite{Forini:2005bs,Vla05,Calcagni:2007wy,Calcagni:2007ef,Mulryne:2008iq}
and applied to several nonlocal systems such as the low-energy limit of string-field theory \cite{Forini:2005bs,Calcagni:2007wy,Calcagni:2007ef,Joukovskaya:2007nq,Joukovskaya:2008zv,Joukovskaya:2008cr,Calcagni:2009tx,Calcagni:2009jb}, NLQG \cite{Calcagni:2018lyd,Calcagni:2018gke,Calcagni:2018fid} and FQG \cite{Calcagni:2025wnn}. The general finding in all these cases was that, if we represent the equations of motion for the fields $\Phi_i$ as the sum of a local operator $\cO_n(\p)=O(\p^n)$ with a certain time-derivative order $n$ plus local interactions $V$ plus a nonlocal operator $\cO_{\rm NL}(\p)$,
\be\label{eq1}
\cO_n^i(\p)\,\Phi_i + V(\{\Phi_i\})+ \Phi_j^p\cO_{\rm NL}^{ij}(\p)\Phi_i^q = 0\,,
\ee
where $p,q=0,1,2,\dots$ are model-dependent fixed powers, then these equations require $n$ initial conditions. This is true \emph{only} for the operators $\cO_{\rm NL}$ considered in the above papers and is \emph{not} valid for an arbitrary form factor $\cO_{\rm NL}$.

In a recent publication \cite{Anselmi:2025uda}, it was shown that systems with fakeons have the same type of Cauchy problem as \Eq{eq1}: the classicized equations of motion have a nonlocal term despite of which one only requires $n=2$ initial conditions to specify the solutions. This conclusion was reached in several toy models with explicit calculations, made possible because the inverse $\cO_{\rm NL}^{-1}$ of the form factor is a local operator in the fakeon case. A general proof was also given in \cite{Anselmi:2025uda}, based on the fact that the inverse of the form factor has an $N$-dimensional kernel, where $N=\deg\cO_{\rm NL}^{-1}(\p)$, which constrains the $N+n$ solutions of the associated higher-derivative system obtained applying $\cO_{\rm NL}^{-1}$ to \Eq{eq1}.

Given the momentum that fakeons are gaining in quantum gravity as a mechanism to preserve unitarity, it is instructive to understand the Cauchy problem of these systems also from the perspective of the diffusion method. There are strong similarities between the classicized fakeon dynamics and the one of fractional field theories, since the former displays integer inverse powers of the d'Alembertian $\B$ \cite{Anselmi:2025uda} while the latter have generic noninteger powers $\B^{\g}$ \cite{Calcagni:2021ipd,Calcagni:2021ljs,Calcagni:2021aap,Calcagni:2022shb}. Since the diffusion method can be adapted to fractional field theories \cite{Calcagni:2025wnn}, the expectation is that it can also work for fakeons. In this paper, we prove just that. Another motivation for this exercise is that the same type of nonlocality $\B^{-n}$ in the fakeonic classicized Lagrangian also appears in cosmological models of infrared nonlocal gravity, which have drawn attention for their impact on the late-time acceleration of the universe and on the gravitational-wave luminosity distance \cite{Barvinsky:2003kg,Deser:2007jk,Belgacem:2017cqo,Belgacem:2020pdz}. The diffusion approach could be useful to find exact or approximate solutions of such models.

In Sec.~\ref{sec4}, we deploy the diffusion method for an interacting two-scalar field theory with a fakeon mode. In Sec.~\ref{diftoy}, we apply the diffusion method to a general class of linear toy models including those of \cite{Anselmi:2025uda}, where we find that the simplification to a linear dynamics actually introduces some technical complications not present in the nonlinear case. Independently of the presence of interactions, we find that the solutions of these systems are fully and unambiguously specified by two initial conditions. Conclusions are in Sec.~\ref{sec5}. 


\section{Fakeons with interactions}\label{sec4}

In this section, we briefly recall scalar fakeon systems with complex and real masses and then apply the diffusion method to both, in a self-contained way.


\subsection{Fakeons with complex masses}\label{dmfac}

Consider a $D=4$ dimensional quantum field theory with fakeons, described by the \emph{interim} Lagrangian 
\ba
\cL&=&-\frac{1}{2}\varphi(\B+m^2)\varphi-V(\varphi)\nn
&& -\frac{1}{2M^{2}}\phi\left[ (\B+m^2)^2+M^4\right]
\phi-\frac{g}{2}\phi\varphi^{2}\,,\qquad\label{inter}
\ea
where $\B=\p_{\mu}\p^{\mu}=\p_t^2-\p_{\bm{x}}^2$ is the d'Alembertian or Laplace--Beltrami operator on Minkowski spacetime with mostly minus signature $(+,-,-,-)$, $m\in\mathbb{R}$ and $M\in\mathbb{R}$ are masses and $g$ is a cross-interaction coupling constant. The field $\phi$ has a quartic kinetic term and can be decomposed in two modes with complex-conjugate squared masses, one of which is a ghost. To remove the ghost and maintain hermiticity of the action, one must remove the whole pair from the spectrum of asymptotic states of the theory, which is done applying the Anselmi--Piva or fakeon prescription \cite{Anselmi:2017yux,Anselmi:2021hab,Anselmi:2025uzj}. According to the latter, one can remove the ghosts by extending momentum integration to complex spatial momenta $\bm{k}$ along certain paths \cite{Anselmi:2017yux,Anselmi:2025uzj} (a simplified calculations in a fixed Lorentz frame can be found in \cite{Anselmi:2025uzj,Liu:2022gun})\footnote{In contrast, in the Lee--Wick--Nakanishi prescription \cite{Lee:1969fy,Lee:1970iw,Lee:1969zze,Nakanishi:1971jj} one keeps $\bm{k}$ real; see also \cite{Modesto:2015ozb,Modesto:2016ofr} for a more recent application of this prescription in higher-derivative quantum gravity.} or, equivalently, by working in Euclidean momenta and then taking the average of the analytic continuations $k^2\to k^2\pm\rmi\e$ of the scattering amplitude around branch cuts \cite{Anselmi:2017yux,Anselmi:2025uzj} (instead of the usual causal Feynman analytic continuation $k^2\to k^2+\rmi\e$). Contrary to other prescriptions applied to complex-conjugate pairs, this procedure preserves both Lorentz invariance and the optical theorem and gives rise to a new but rigorous diagrammar \cite{Anselmi:2021hab}. The net result is that the offending modes are projected out of the spectrum order by order in perturbation theory and can only appear as purely virtual particles (the so-called fake particles or \emph{fakeons}) in the internal loops of scattering amplitudes. At the leading perturbative order, this is achieved by plugging in the equation of motion $\de\cL/\de\phi=-[(\B+m^2)^2+M^4]\phi-M^2 g\vp^2/2=0$ back into the Lagrangian \Eq{inter} or, equivalently, integrating out the fakeon field $\phi$ in the path integral. Then, one obtains the
nonlocal \emph{classicized} Lagrangian \cite{Anselmi:2025uda}
\ba
\cL_{\text{cl}}&=&-\frac{1}{2}\varphi(\B+m^2)\varphi-V(\varphi)\nn
&&+\frac{g^{2}}{8}\varphi ^{2}
\left. \frac{M^2}{(\Box+m^2)^2+M^4}\right|_{\textrm{f}}\vp^2\,,\label{milan}
\ea
which governs the classical dynamics of $\vp$ at leading order in the coupling $g$.
The $O(g^2)$ nonlocal term is evaluated with the fakeon (``f'') prescription and is characterized by complex-conjugate poles with masses $\pm\rmi M^2$. The \emph{interim} Lagrangian \Eq{inter} is fourth-order in spacetime derivatives and has four initial conditions $\vp(t_0,\bm{x})$, $\dot\vp(t_0,\bm{x})$, $\phi(t_0,\bm{x})$ and $\dot\phi(t_0,\bm{x})$. However, this is not the correct counting for the classical system, which should be done on the classicized Lagrangian \Eq{milan} and its associated equation of motion\footnote{Equations motion in nonlocal systems are calculated with the usual variational principle, representing the nonlocal operators $\g(\B)$ in a suitable way \cite{BasiBeneito:2022wux,Calcagni:2025wnn,Pais:1950za,Calcagni:2007ru}. 
The series representation $\g(\B)=\sum_n a_n \B^n$ is always possible, even in cases where it is only formal and some of the coefficients $a_n$ need to be regularized [e.g., $\g(\B)=\sqrt{\B}\to\sqrt{\B+\ve^2}\simeq \ve+\B/(2\ve)+\dots$], since all the terms are fully resummed back to $\g(\B)$ (when varying with respect to matter fields) or its functional derivatives (when varying with respect to the metric) \cite{Calcagni:2005xc,Koshelev:2013lfm,Biswas:2013cha,Conroy:2014eja}. Alternative representations may work just fine and yield the same result \cite{Calcagni:2018lyd}. For instance, as we see below, using the natural representation \Eq{bkrep} one can obtain \Eq{milan} from \Eq{inter} by writing them as \Eq{milan3} and \Eq{eomph}, respectively. Once a representation is selected, the functional space upon which the form factor $\g(\B)$ acts is chosen accordingly.} 
\be\label{milan2}
(\B+m^2)\vp +V'(\vp)-\frac{g^2}{2}\vp\frac{M^{2}}{(\B+m^2)^2+M^4}\vp^2=0\,.
\ee
This equation is nonlocal and has an infinite number of derivatives, thus leading one to wonder about the infamous paradox \cite{Eliezer:1989cr,Moeller:2002vx} that knowing the infinitely many initial conditions $\vp^{(n)}(t_0,\bm{x})$ of a nonlocal system is tantamount to already knowing the Taylor-expanded solution $\vp(x)=\sum_{n=0}^{+\infty}\vp^{(n)}(t_0,\bm{x})(t-t_0)^n/n!$ beforehand. However, since the fakeon $\phi$ does not contribute directly to the classical dynamics \Eq{milan2} of physical degrees of freedom, intuitively the number of initial conditions should be two, as one would obtain by naively setting $\phi=0$ in the \emph{interim} Lagrangian \Eq{inter}.

To check this counting, we solve the Cauchy problem with the diffusion method. With this in mind, we first represent the nonlocal form factor in a way compatible with the Anselmi--Piva prescription. We use the fact that the inverse of a generic positive operator $A$ can be written in terms of the well-known Balakrishnan--Komatsu integral representation \cite{Calcagni:2025wnn}
\be\label{bkrep}
\frac{1}{A}=\int_0^{+\infty}\rmd \tau\,\rme^{-\tau A},\qquad \Re\,A>0\,,
\ee
where ``$\Re\,A>0$'' is understood to be applied to the spectrum of $A$. Plugging \Eq{bkrep} into \Eq{milan} with
\be
A=1+\lst^4(\B+m^2)^2\,,\qquad \lst\coloneqq \frac{1}{M}\,,
\ee
we can express the classicized Lagrangian as a nonlocal system with a type of exponential nonlocality that we know how to treat with the diffusion method:
\ba
\cL_{\text{cl}}&=&-\frac{1}{2}\vp(\B+m^2)\vp-V(\vp)\nn
&&+\frac{g^2\lst^2}{8}
\int_0^{+\infty}\rmd \tau\,\rme^{-\tau}\vp^2\,\rme^{-\tau \lst^4(\B+m^2)^2}\vp^2\,,\qquad\label{milan3}
\ea
where $\tau$ is a dimensionless parameter. The equation of motion produced by this Lagrangian is \Eq{milan2}
\ba
0&=&(\B+m^2)\vp+V'(\vp)\nn
&&-\frac{g^2\lst^2}{2}
\int_0^{+\infty}\rmd \tau\,\rme^{-\tau}\vp\,\rme^{-\tau \lst^4(\B+m^2)^2}\vp^2\,.\qquad\label{eomph}
\ea
The local limit of this equation is towards a zero length scale $\lst\to 0$, i.e., where the fakeon has infinite mass $M\to+\infty$. We call $\vp_0(x)$ the solution of the equation of motion in this local limit,
\be\label{eomph0}
\lst\to 0\,:\qquad (\B+m^2)\vp_0+V'(\vp_0)=0\,.
\ee


\subsection{Fakeons with real masses}\label{dmfac2}

The classicized Lagrangian of this system \cite{Anselmi:2025uda} can be rewritten as the limit $M\equiv\ve\to 0$ of \Eq{milan2} augmented by a $(\B+m^2)/M^2$ operator,
\ba
\cL_{\text{cl}}&=&-\frac{1}{2}\vp(\B+m^2)\vp-V(\vp)\nn
&&+\lim_{\ve\to 0}\frac{g^2}{8}\,\varphi^{2}\frac{\B+m^2}{(\B+m^2)^2+m^4\ve^2}\,\varphi^{2}\nn
&=&-\frac{1}{2}\vp\B\vp-\frac{1}{2\lst^2}\varphi^{2}-V(\vp)\nn
&&+\lim_{\ve\to 0}\frac{g^2\lst^2}{8}\,\varphi^{2}\frac{\lst^2\B+1}{(\lst^2\B+1)^2+\ve^2}\,\varphi^{2}\,,\qquad\label{milanreal}
\ea
where $\lst=m^{-1}$. Due to the factor $\lst^2\B+1$ in the numerator, the nonlocal operator in the quartic term is not positive. However, we can apply exactly the same representation \Eq{bkrep},
\ba
\cL_{\text{cl}}&=&-\frac{1}{2}\vp\B\vp-\frac{1}{2\lst^2}\varphi^{2}-V(\vp)\nn
&&+\lim_{\ve\to 0}\frac{g^2\lst^2}{8}
\int_0^{+\infty}\rmd \tau\,\rme^{-\tau \ve^2}\nn
&&\qquad\qquad\quad\times\vp^2(\lst^2\B+1)\,\rme^{-\tau (\lst^2\B+1)^2}\vp^2\,.\qquad\label{milan3real}
\ea


\subsection{Diffusion method}

The diffusion method is a systematic recipe that helps to count the number of initial conditions in certain nonlocal classical systems and to find exact or approximate solutions. Its logic is the following:
\begin{enumerate}
\item Starting from a $D$-dimensional nonlocal system, adding an extra artificial coordinate $r$ one promotes all fields to a system living in $D+1$ dimensions.
\item The action of nonlocal operators is represented as a certain operation (typically, a translation, possibly integrated over) on $r$, so that the $D+1$ system is local in spacetime coordinates (i.e., is of finite derivative order) but nonlocal in $r$. We call this system \emph{localized}.
\item Fixing one of the $(D+1)$-dimensional fields to zero, each of the other fields obey a diffusion equation where $r$ is diffusion time; this equation can be solved specifying a finite number of initial conditions in the spacetime time coordinate $t=x^0$. In other words, the original Cauchy problem is transformed into a boundary problem and the infinite number of initial conditions in the nonlocal system are traded for a finite number of initial conditions on a field with more directions \cite{Mulryne:2008iq,Calcagni:2018lyd}.
\item While for general $r$ the solutions of the diffusion equations are not guaranteed to be solutions of the other equations of the $D+1$ system, one postulates the existence of a certain slice $r=r_*$ where they obey the same equations of motion as the fields in the nonlocal system. 
\end{enumerate}
Item 4 is commented upon later and is modified in the less general setting of Sec.~\ref{diftoy}.

Let us apply this recipe to the $D=4$ fakeons with complex masses of Sec.~\ref{dmfac} (the case with real masses of Sec.~\ref{dmfac2} is dealt with at the end) and consider \emph{ex novo} a system local in spacetime directions with three scalar fields $\Phi$, $\chi$ and $\la$ living in $4+1$ dimensions:
\bs\label{locsy}\ba
\cS&=&\int\rmd^4x\,\rmd r\,(\cL_\Phi+\cL_\chi+\cL_{\la})\,,\label{locsy1}\\
\cL_\Phi &=& -\frac{1}{2}\Phi(r,x)\B\Phi(r,x)-\frac{m^2}{2}\Phi^2(r,x)-V[\Phi(r,x)]\nn
&&+\frac{g^2\lst^2}{8}\,f(r)\int_0^{+\infty}\rmd \tau\,\rme^{-\tau}\Phi^2(r,x)\Phi^2(r+\tau,x)\,,\nn\label{locsy2}\\
\cL_\chi &=& -\frac{g^2\lst^2}{8}\int_0^{+\infty}\rmd\tau\,\rme^{-\tau}f(r-\tau)\nn
&&\times\int_0^{\tau} \rmd q\,\chi(r-q,x)[\p_{r'}+\lst^4(\B+m^2)^2]\Phi^2(r',x)\,,\nn\label{locch2}\\
\cL_{\la} &=& \la(r,x)\left[\chi(r,x)-\Phi^2(r,x)\right],\label{locsy4}
\ea\es
where $r\geq 0$ is an extra direction with dimensionality $[r]=0$, $r'=r+q-\tau$, $\chi$ and $\la$ are two auxiliary fields with $[\chi]=[\la]=2$ and $f(r)$ is a smooth function of $r$ to be determined later. 

The equation of motion $\de\cS/\de\chi=0$ yields (hiding the spacetime coordinate $x$)
\ba
\lambda(r)&=& \frac{g^2\lst^2}{8}\int_0^{r_*}\rmd\tau\,\rme^{-\tau}\!\!\int_0^{\tau} \rmd \rho\,f(\rho+r-\tau)\nn
&&\times[\p_{r}+\lst^4(\B+m^2)^2]\Phi^2(r+2\rho-\tau)\,.\qquad\label{tempo}
\ea
We are not interested in all the solutions of this extended system and we can select a field configuration such that $\Phi^2$ diffuses. The choice 
\be
\la(r,x)=0\label{la0}
\ee
implies
\ba\label{difeq}
\left[\p_{r}+\lst^4(\B+m^2)^2\right]\Phi^2(r,x)=0\,.
\ea
One can check \Eq{difeq} with a direct derivation from the action but a quicker proof of this is as follows. Multiply \Eq{difeq} by $\delta(r-r^\prime)$ and integrate in $\rmd r$. Then shift $r$ by $-2q+\tau$. At that point, the $q$ integral restricts the $\tau$ integral, which is straightforward, to $\tau\geq |r-r^\prime|$. After the integral on $\tau$, we get the convolution
\ba
0&=&g^4\lst^4\int_{-\infty}^{+\infty}\rmd r\, \rme^{-|r-r^\prime|}G(r)\,,\nn
G(r)&=& [\p_{r}+\lst^4(\B+m^2)^2]\Phi^2(r)\,,
\ea
which is easily Fourier transformed to conclude that $G(r)=0$. Before proceeding, we comment on the fact that \Eqq{tempo} admits nontrivial profiles $\la(r)\neq 0$.  Equation \Eq{la0} is equivalent to look for special solutions of the localized system $\cS$, namely, only solutions obeying the diffusion equation. $\cS$ might also have other solutions but these do not correspond to, or at least they do not help to find, solutions of the nonlocal system. If $\la\neq 0$, the diffusion equation is no longer valid and one cannot write the action of a nonlocal operator as a translation in the extra coordinate $r$. Thus, the correspondence between the localized and the nonlocal system is not one-to-one and the degrees of freedom of the latter are a projection of the degrees of freedom of the former. 

The equation of motion of $\la$ gives
\be\label{constrl}
\chi(r,x)=\Phi^2(r,x)\,,
\ee
which implies that also $\chi$ obeys the diffusion equation \Eq{difeq}:
\ba\label{difeqhi}
\left[\p_{r}+\lst^4(\B+m^2)^2\right]\chi(r,x)=0\,.
\ea
Finally, to compute $\de\cS/\de\Phi=0$ we note that 
\ban
\cL_{\chi}&=&\frac{g^2\lst^2}{8}\int_0^{+\infty}\rmd\tau\,\rme^{-\tau}f(r-\tau)\nn
&&\times\int_0^{\tau}\rmd q\Big\{-\p_{q}[\chi(r-q)\Phi^2(r')]\nn
&&\qquad+\Phi^2(r')[\p_{q}-\lst^4(\B+m^2)^2]\chi(r-q)\Big\}\nn
&=&\frac{g^2\lst^2}{8}\int_0^{+\infty}\rmd\tau\,\rme^{-\tau}f(r-\tau)\nn
&&\quad\times\left\{[\chi(r)\Phi^2(r-\tau)-\chi(r-\tau)\Phi^2(r)]\vphantom{\int_0^\tau}\right.\nn
&&\quad\left.+\int_0^{\tau} \rmd q\,\Phi^2(r')[\p_{r'}-\lst^4(\B+m^2)^2]\chi(r-q)\right\},
\ean
and the equation of motion of $\Phi$ is
\ban
0&=&(\B+m^2)\Phi(\bar r)+V'[\Phi(\bar r)]+2\la(\bar r)\,\Phi(\bar r)\nn
&&-\frac{g^2\lst^2}{4}
\int_0^{+\infty}\rmd \tau\,\rme^{-\tau}\Phi(\bar r)\nn
&&\times\left[\vphantom{\int}f(\bar r)\Phi^2(\bar r+\tau)+f(\bar r)\chi(\bar r+\tau)\right.\nn
&&\qquad+f(\bar r-\tau)\Phi^2(\bar r-\tau)-f(\bar r-\tau)\chi(\bar r-\tau)\nn
&&\left.\qquad-\int_{\bar r}^{\bar r+\tau} \rmd r\,[\p_{-\bar r}+\lst^4(\B+m^2)^2]\chi(2r-\bar r-\tau)\right].
\ean
In all these expressions, we assumed that the supports of the deltas of the functional variations $\de\Phi(r)/\de\Phi(\bar r)=\de(r-\bar r)$ lie in the semi-interval $r\in(0,+\infty)$. 
Imposing \Eq{la0} and \Eq{constrl}, this equation becomes
\ba
0&=&(\B+m^2)\Phi(r,x)+V'[\Phi(r,x)]\nn
&&-\frac{g^2\lst^2f(r)}{2}\Phi(r,x)
\int_0^{+\infty}\rmd \tau\,\rme^{-\tau}\Phi^2(r+\tau,x)\,.\nn\label{eom1}
\ea


We can get rid of the nonlocal factor in the equations of motion in order to recast the system as a complicated $O(\B^3)$ equation in terms of the field $\chi=\Phi^2$, whose solutions automatically solve also the diffusion equation. 
 This higher-derivative parent equation, which we do not show here for this nonlinear system, turns out to be very useful to solve linear models such as those of Sec.~\ref{diftoy}, where we will employ it extensively.
 Here, we limit ourselves to note that the system is made of two independent local equations, the diffusion equation \Eq{difeqhi} of derivative order $2N_{\rm diff}=4$ and the parent higher-derivative equation of order $2N=6$. Therefore, the number of initial conditions is $\cN= 6$ but for the original nonlocal system this is an overcounting that can be addressed as follows.

In the diffusion method, the diffusion process describes how a field evolves in the $r$-direction from the initial value\footnote{This value is a convention and what matters is that $r$ be finite at the starting point and that evolution take place for growing $r$ (one-way directionality of diffusion). It is easy to show that diffusion does not work as a boundary problem at infinity starting from $r=\infty$.} $r=0$ to some final value $r_*$ which we set to be finite at $r=r_*=1$ here. At the end of the flow, the dynamics of the localized system \Eq{locsy} must coincide by construction with the one of the nonlocal system \Eq{milan3}, i.e., \Eq{eom1} must reproduce \Eq{milan2}, which happens only if
\be\label{flowend}
\vp(x)=\Phi(1,x)\,,\qquad f(1)=1\,. 
\ee
The solution of the diffusion equation requires one to know the system on the $r=0$ slice but, to do so, the dynamics must necessarily become local on this slice, since otherwise we would be back to a nonlinear nonlocal dynamics for which we have no systematic methods at hand to solve \cite{Calcagni:2018lyd,Calcagni:2018gke,Calcagni:2025wnn}. In the present case, locality on the $r=0$ slice for general solutions is achieved by fixing $f(r)$ so that $f(0)=0$. 

To solve the localized system \Eq{locsy}, we look for completely general solutions $\Phi(r,x)$ and regard the diffusion along $r$ as a renormalization-group flow. In this scenario, the diffusion parameter $r$ acts as a renormalization-group scale \cite{Calcagni:2018gke} always coupled to the length scale in the dimensionful hypervolume quantity $\varrho\coloneqq r\lst^4$.\footnote{The system \Eq{locsy} is all written in terms of the combination $\varrho$ except in the diffusion term $\lst^4(\B+m^2)$. This is only a choice of conventions in order to have a diffusion equation with linear diffusion direction. Imposing an $r\lst^4(\B+m^2)$ term does not change the final result below.} The $r\to 0$ limit corresponds to a zero $\varrho$, while the limit $r\to 1$ corresponds to the physical hypervolume scale $\varrho=\lst^4$ of the nonlocal system. This suggests to fix $f(r)$ in \Eq{locsy2} as
\be\label{fr}
f(r) =\sqrt{r}\, h(r)\,,
\ee
where $h(r)$ is a generic smooth function of $r$ such that $h(1)=1$. Without any loss of generality, we can also set $h(r)\equiv 1$.

The function \Eq{fr} is crucial to have, at the beginning of the diffusion flow on the $r=0$ slice, a local limit of \Eq{eom1} that coincides exactly with the $\lst\to 0$ limit \Eq{eomph0}:
\be\label{eomph02}
r\to 0\,:\qquad (\B+m^2)\Phi(0,x)+V'[\Phi(0,x)]=0\,.
\ee
In the toy models of Sec.~\ref{diftoy}, $m=0$ and $V'=0$ and the local equation of motion is $\B\Phi(0,x)=0$; this will require a different handling of the diffusion flow. 

On the other hand, at the end of the diffusion flow at $r=1$, \Eqq{eom1} corresponds to the equation of motion \Eq{eomph} if we make the identification \Eq{flowend}, so that
\ba
0&=&(\B+m^2)\Phi(1,x)+V'[\Phi(1,x)]\nn
&&-\frac{g^2\lst^2}{2}\Phi(1,x)
\int_0^{+\infty}\rmd \tau\,\rme^{-\tau}\Phi^2(1+\tau,x)\,.\qquad\label{eom1r1}
\ea
The general solution $\Phi(0,x)=\vp_0(t,\bm{x})$ of \Eq{eomph02} needs to uniquely specify two initial conditions at the initial time $t=t_0$:
\be\label{inco}
\Phi(0,t_0,\bm{x})\,,\qquad \dot\Phi(0,t_0,\bm{x})\,.
\ee
These initial conditions are also necessary and sufficient to find the general solution $\Phi(r,x)$ of the diffusion equation \Eq{difeq}. In other words, the initial conditions \Eq{inco} determine $\Phi(0,x)$ uniquely and are enough to solve the system in cascade: 
\begin{enumerate}
\item Via \Eq{inco}, one solves \Eq{eomph02} and finds $\Phi(0,x)=\vp_0(x)$.
\item With this initial field configuration, one solves \Eq{difeq} and finds $\Phi(r,x)$.
\item Finally, one must check that the function $\Phi(1,x)$ is also a solution of \Eq{eom1r1}.
\end{enumerate}
The last step in the list is highly nontrivial and may be impossible in general, where one must content oneself to find asymptotic nonlocal solutions (for exact and asymptotic diffusing solutions in other nonlocal systems, see \cite{Calcagni:2007ef} and \cite{Calcagni:2007wy,Calcagni:2007ru,Calcagni:2008nm,Calcagni:2009tx,Calcagni:2009dg,Calcagni:2009jb}, respectively). 

The net result is that 
 one only needs $\cN=2$ initial conditions. Due to technical complications inherent to the linear dynamics, in Sec.~\ref{diftoy} we use a different way (which we could apply here as well) to obtain a local equation of motion in the initial slice $r=0$.
 

To summarize, independently of the form of the potential $V(\vp)$, the diffusion method specifies uniquely that the number of initial conditions necessary to find solutions of the nonlocal system \Eq{milan2} is equal to 2. These results generalize to fakeon models the conclusion of \cite{Calcagni:2025wnn} that the number of initial conditions is two for each physical field of any nonlocal system whose associated local system (i.e., in the limit of small nonlocality length scale $\lst$) is second-order in time derivatives:
\begin{theorem*}
{The number of initial conditions is $\cN=2$ for each physical field in any theory with fakeons whose classicized Lagrangian is second-order in time derivatives when nonlocality is switched off.}
\end{theorem*}

All the above can be adapted to fakeons with real masses (Sec.~\ref{dmfac2}). The diffusion method applies in the same way, with \Eq{locsy2} and \Eq{locch2} having an extra operator $\lst^2\B+1$ inside the integrals. The number of initial conditions is two as in all the other models in this paper.


\section{Diffusion method for linear toy models}\label{diftoy}

In Sec.~\ref{sec4}, we used the full version of the diffusion method for a system with nonlinear interactions. In the absence of interactions, the fourth item in the list defining the method must be modified accordingly:
\begin{enumerate}
\item[4.] For the toy models discussed in the present section, a simplified version of the method holds and the solutions of the diffusion equation are also solutions of the $(D+1)$ system for all $r$.
\end{enumerate}

We begin by making a general formulation of all the linear examples of \cite{Anselmi:2025uda}. Consider the nonlocal $D=1$ mechanical system
\bs\label{toy2}\ba
\cL &=& -\frac12\, x(t)\,\p_t^2x(t)-\frac12\,\om^2 x(t) F(\p_t^2)\, x(t)\,,\qquad\\
F&=&\frac{\cB(\p_t^2)}{\cA(\p_t^2)}\,,
\ea\es
with equation of motion 
\be\label{toy2eom}
\p_t^2x(t)+\om^2 F(\p_t^2)\, x(t)=0\,,
\ee
where $\om\in\mathbb{R}$ and in all our examples $\cA,\,\cB$ are polynomials of $\p_t^2$ of degree ${\rm deg}\,\cA\geq {\rm deg}\,\cB$ such that $\cB(0)=1$. Hence, $\cA$ (and, similarly, $\cB$) is a local operator of derivative order ${\rm ord}\,\cA=2\,{\rm deg}\,\cA$. More general $\cA$ and $\cB$ with odd powers of $\p_t$ are also possible and they do not modify much the following discussion.
The classicized Lagrangian \Eq{milan} is akin to a toy model with $\cA=O(\p_t^4)$ and $\cB=1$ but with a nonlinear dependence on the field in the nonlocal term. 

We assume that $\cA$ is a non-negative operator, meaning that, if the spectrum of $\p_t^2$ is $\Om_i^2$ for some complex $\Om_i$ and $i=1,2,\dots$, then
\be\label{posiA}
\Re\,\cA(-\Om_i^2)>0\,,\qquad \Om_i\in\mathbb{C}\,.
\ee
Under this condition, we can apply the Balakrishnan--Komatsu representation \Eq{bkrep} and write $F$ as 
\be\label{posiF}
F(\p_t^2)
=\int_0^{+\infty}\rmd\t\,\cB(\p_t^2)\,\rme^{-\t \cA(\p_t^2)}\,.
\ee

Linear nonlocal systems can be solved easily via a field redefinition \cite{Pais:1950za,Barnaby:2007ve}. In the case of \Eq{toy2}, calling $\tilde x\coloneqq \sqrt{1+\om^2 F(\p_t^2)/\p_t^2}\,x$ and integrating by parts, we get $\cL = -(1/2)\, \tilde x(t)\p_t^2\tilde x(t)$, a canonical system with two initial conditions. However, depending on the pole structure of the factor $F$, this field redefinition might artificially remove some degrees of freedom of the original system. As shown in \cite{Anselmi:2025uda}, this problem does not arise in fakeonic systems, even if $F$ has indeed complex poles, and the number of initial conditions required to specify the general solution is two. Below, we apply the diffusion method to this case not only as an independent confirmation of \cite{Anselmi:2025uda} but also to better understand how it works for nonlocal operators of the form considered in this paper, which have various applications mentioned in Sec.~\ref{sec5}.



\subsection{Active versus passive slicing}

The construction of a diffusing localized system with a field $X(r,t)$ allowing us to solve the original nonlocal system \Eq{toy2eom} is hindered by a problem not present in the interacting case of the previous section. To understand this, consider the diffusion equation
\be\label{difx}
\left[\p_r+\cA(\p_t^2)\right]X(r,t)=0
\ee
and the localized equation of motion (local in $t$ but nonlocal in $r$) analog to \Eq{eom1} and equivalent to \Eq{toy2eom} on a certain $r$-slice:
\be\label{eom2f}
\p_t^2X(r,t)+\om^2f(r)\int_0^{+\infty}\rmd\t\,\cB(\p_t^2) X(r+\t,t)=0\,.
\ee
It is not difficult to prove that this equation can be derived from a localized action as we did in the previous section. Noting that
\ban
&&\p_r\left[\int_0^{+\infty}\rmd\t\,\cB\,X(r+\t,t)\right]\\
&&\qquad \stackrel{\t'=\t+r}{=} \p_r\left[\int_r^{+\infty}\rmd\t'\,\cB\,X(\t',t)\right]=-\cB\,X(r,t)\,,
\ean
applying the operator $\p_r$ onto \Eq{eom2f} and using the above formula and the diffusion equation \Eq{difx}, we get
\ben
\left[\left(\cA+\frac{\p_r f}{f}\right)\p_t^2+\om^2f\cB\right]X(r,t)=0\,.
\een
 If there are no repeated roots (as is the case in our toy models), then the most general solution of this higher-derivative expression is 
\ba
&& X(r,t)=\sum_{i=1}^N \left[p_i^+(r)\,\rme^{\rmi\Om_i(r) t}+p_i^-(r)\,\rme^{-\rmi\Om_i(r) t}\right],\nn
&& \Om_i(r)\in\mathbb{C}\,,\label{solgen}
\ea
where $N=1+{\rm ord}\,\cA$ (ord is the derivative order in $\p_t$) and 
$\Om_i(r)\in\mathbb{C}$ are the roots of the $r$-dependent characteristic equation
\ben
\left[\cA_i+\frac{\p_r f}{f}\right]\Om_i^2-\om^2f\cB_i=0\,,
\een
where
\be\label{cai}
\cA_i\coloneqq \cA(-\Om_i^2)\,,\qquad \cB_i\coloneqq \cB(-\Om_i^2)\,.
\ee
However, plugging \Eq{solgen} into the diffusion equation \Eq{difx} yields $\p_r p_i^\pm/p_i^\pm+\cA\pm\rmi t \p_r\Om_i=0$ for each $i$. Therefore, $p^+_i\propto p^-_i$ and $\Om_i={\rm const}$, which implies $f(r)\equiv 1$.

This conclusion forbids us to follow the recipe of Sec.~\ref{sec4} and to start the diffusion process from a slice with the nonlocal term completely switched off. This problem can be overcome by defining the localized system in such a way as to switch off the nonlocal term in a certain $r$-limit. While for a nonlinear system we would solve the diffusion equation with some initial condition $X(0,t)$ obeying a local equation of motion and then make the localized system dynamically equivalent to the original nonlocal one at some slice $r=r_*$ where 
\be\label{x1}
x(t)=\lim_{r\to r_*} X(r,t)
\ee
is the general solution of \Eq{toy2eom}, in the linear case we evaluate the dynamics on the $r=r_*$ slice and then vary $r_*$ from 0 to some end point we will set below to be at infinity:
\be\label{x2}
x(t)=\lim_{r_*\to+\infty} X(r_*,t)\,.
\ee
This is somewhat analogous to the difference between a passive and an active symmetry transformation. In the case of \Eq{x1}, we follow the diffusing dynamics from the initial slice $r=0$ to some final slice $r=r_*$; this is an active view where a given $r$-slice is made moving along the $r$-direction. In contrast, in the case of \Eq{x2} we focus on a given slice $r=r_*$ and let this value of $r$ evolve; this is a passive view where the slice is fixed and, roughly speaking, it is the $r$-direction to be shifted through it. 

From the point of view of the physics of \Eq{toy2eom}, it does not matter whether one adopts an active or a passive view, as long as the system is self-consistent. In Sec.~\ref{sec4}, we chose the active view because it is commonly used in other applications of the diffusion method but, here, we have to employ the passive view because the other meets an obstacle. The final result, i.e., the individuation of the solutions of the nonlocal system \Eq{toy2eom}, must be the same regardless of the view adopted as well as of the way one implements any such view.

There are various forms of building a localized system in the passive view \Eq{x2}. For example, instead of \Eq{bkrep} we can consider the Balakrishnan--Komatsu representation of the operator
\ban
\frac{1}{A}&\coloneqq& \lim_{r_*\to+\infty}\frac{1-\rme^{-r_* A}}{A}\\
&=&\lim_{r_*\to+\infty}\int_0^{r_*}\rmd \tau\,\rme^{-\tau A},\qquad \Re\,A>0\,,
\ean
where $r_*>0$ is a sort of cutoff or regulator which is eventually removed at the end of the calculation. The crucial difference with respect to \Eq{bkrep} is that the operator $(1-\rme^{-r_* A})/A=r_*+O(A^2)$ is entire, so that $F$ can be written as the limiting case of an entire operator. We checked that it is possible to construct a localized action such that \Eqq{eom2f} is replaced by
\be\label{eom2rst}
\p_t^2X(r,t)+\om^2\int_0^{r_*}\rmd\t\,\cB(\p_t^2) X(r+\t,t)=0\,,
\ee
which reduces to the local equation
\be\label{inicon}
\p_t^2X(0,t)=0\qquad \Longrightarrow\qquad X(0,t)=a_0+a_1 t\,\,\,\,
\ee
in the limit $r=r_*\to 0$, while recovering \Eq{toy2eom} with $f=1$ in the limit $r_*\to+\infty$ using \Eq{x2}. The general solution of the dynamics \Eq{difx} and \Eq{eom2rst} is a linear superposition of infinitely many exponentials like \Eq{solgen} with $r_*$-dependent frequencies $\Om_i$ obeying a transcendental characteristic equation. The solutions of this equation can be studied analytically but here we pursue a much simpler alternative eventually giving the same final result [the solution \Eq{physol}], where the cutoff $r_*$ appears not as a regulator in the $\t$ integral but as a coupling strength parameter attached to $\om^2$:
\be
\om^2\to \om^2 f(r_*)\,,
\ee
where $f(r_*)$ must be such that $f(0)=0$ and $f(+\infty)=1$. An example of profile, as good as any with the above boundary conditions, is
\be\label{fexp}
f(r_*)=1-\rme^{-r_*}\,.
\ee


\subsection{Localized system}

We hereby propose to localize \Eq{toy2} with the system
\bs\label{locsyL2}\ba
\cS&=&\int\rmd t\,\rmd r\,(\cL_X+\cL_Y+\cL_{\la})\,,\\
\cL_X &=& -\frac12\, X(r,t)\p_t^2X(r,t)\nn
&&-\frac12\,\om^2f(r_*) \int_0^{+\infty}\rmd\t\, X(r,t) \cB(\p_t^2) X(r+\t,t)\,,\nn\\
\cL_Y &=&\frac{\om^2f(r_*)}{2} \int_0^{+\infty}\rmd\t \int_0^{\t}\rmd q\,Y(r-q,t)\nn
&&\qquad\times\left[\p_{q}+\cA(\p_t^2)\right]X(r+q-\t,t)\,,\\
\cL_\la &=& \la(r)[Y(r,t)-\cB(\p_t^2)X(r,t)]\,,
\ea\es
where $r$ is a fictitious extra ``direction,'' $r,q,\t$ are dimensionless and $r_*$ is a regularization parameter which will be sent to infinity at the end of the calculation. The function $f$ can be taken to be \Eq{fexp} or any other with the conditions $f(0)=0$ and $f(+\infty)=1$.

The equation of motion $\de\cS/\de\la=0$ is
\be
Y(r,t)=\cB(\p_t^2)X(r,t)\,,
\ee
while $\de\cS/\de Y=0$ for $\la=0$ gives the diffusion equation \Eq{difx}. Following similar steps as in Sec.~\ref{dmfac}, the equation of motion $\de\cS/\de X=0$ for $\la=0$ is
\be\label{eom2}
\p_t^2X(r,t)+\om^2f(r_*)\int_0^{+\infty}\rmd\t\,\cB(\p_t^2) X(r+\t,t)=0\,.
\ee


\subsection{Higher-derivative dynamics}\label{locdy}

For the class of models considered in this paper, it is possible to recast the nonlocal dynamics as a higher-derivative dynamics with a larger space of solutions. Start from the expression
\ba
&&\p_r\left[\int_0^{+\infty}\rmd\t\,\cB\,X(r+\t,t)\right]\nn
&&\qquad \stackrel{\t'=\t+r}{=} \p_r\left[\int_r^{+\infty}\rmd\t'\,\cB\,X(\t',t)\right]\nn
&&\qquad\quad=-\cB\,X(r,t)\,.\label{useful1}
\ea
Applying the operator $\p_r$ onto \Eq{eom2}, noting that it commutes with $\p_t^2$ and using \Eq{difx} and \Eq{useful1}, we get
\be\label{masterloc}
\left[\cA\p_t^2+\om^2f(r_*)\cB\right] X(r,t)=0\,.
\ee
Since both $\cA$ and $\cB$ are polynomials of $\p_t^2$ with $\deg\,\cA\geq\deg\,\cB$, ${\rm ord}\,\cA\geq {\rm ord}\,\cB$ (deg is the degree of the polynomial in $\p_t^2$, while ord is the derivative order in $\p_t$), equation \Eq{masterloc} is higher-derivative of some order $2N=2+{\rm ord}\,\cA$. From the nonlocal nature of the form factor in \Eq{toy2} and its representation in \Eq{posiF}, it is clear that the derivative order of the diffusion equation is $2N_{\rm diff}={\rm ord}\,\cA=2\deg\,\cA$, so that $N=1+N_{\rm diff}$.

The most general solution of \Eq{masterloc} is a linear superposition of exponentials with $r_*$-dependent frequencies:
\ban
X_{\textsc{hd}}(r,t)&=&\sum_{i=1}^N \left[p_i^+(r,r_*,t)\,\rme^{\rmi\Om_i(r_*)t}\right.\\
&&\qquad\left.+p_i^-(r,r_*,t)\,\rme^{-\rmi\Om_i(r_*)t}\right],
\ean
where HD stands for higher-derivative and $p_j^\pm$ are polynomials in $t$ of degree $m_i-1$ with $(r,r_*)$-dependent coefficients, where $m_j$ is the multiplicity of the $i$th root $\Om_i\in\mathbb{C}$ of the characteristic equation associated with \Eq{masterloc}. In particular, if there are no repeated roots then $m_i=1$ and $p_i^\pm(r,r_*,t)=p_i^\pm(r,r_*)$ for all $i$; such is actually the case in our toy models. Compatibility with the diffusion equation \Eq{difx} requires $p_i^\pm(r,r_*)=c_i^\pm(r_*)\,\exp\{-r\cA[-\Om_i^2(r_*)]\}$:
\ba
X_{\textsc{hd}}(r,t)&=&\sum_{i=1}^N \rme^{-r\cA_i}\nn
&&\times\left[c_i^+(r_*)\,\rme^{\rmi\Om_i(r_*) t}+c_i^-(r_*)\,\rme^{-\rmi\Om_i(r_*) t}\right],\nn
\Om_i\in\mathbb{C}\,,\label{solgenfin}
\ea
where $\cA_i$ is defined in \Eq{cai}, $c_i^\pm(r_*)$ are arbitrary cutoff-dependent constants and $\Om_i^2(r_*)$ are the $N$ roots of the characteristic equation ($r_*$-dependence omitted everywhere)
\be\label{masterlainf}
\cC_{\textsc{hd}}(\Om_i^2)\coloneqq \cA(-\Om_i^2)\Om_i^2-\om^2f\cB(-\Om_i^2)=0\qquad \forall\,i\,.
\ee

Thus, we have found that the localized system under study only admits \emph{stationary} solutions \Eq{solgenfin}, i.e., which diffuse trivially according to a factorized dependence on the $r$-coordinate \cite{Calcagni:2007ru,Calcagni:2010ab}. In other nonlocal systems with different form factors, there do exist nontrivial nonstationary solutions, both asymptotic \cite{Calcagni:2007wy,Calcagni:2007ru,Calcagni:2008nm,Calcagni:2009tx,Calcagni:2009dg,Calcagni:2009jb} and exact \cite{Calcagni:2007ef}.

Equation \Eq{solgenfin} is the superposition of $N=1+\deg\,\cA$ solutions of the characteristic polynomial \Eq{masterlainf} and these are also solutions of the localized equation of motion \Eq{eom2}. In this case, the number of initial conditions is $\geq 2$ but $\leq 2N$ [depending on whether some roots violate \Eq{posiA}], no (or only some) degree of freedom is projected out of the spectrum and the system has, in general, unstable modes. The number $N_-$ of roots $\Om_i$ violating the condition \Eq{posiA} ($\Re\,\cA_i>0$) can be computed by the argument principle:
\ba
N_- &=&\frac{1}{2\pi\rmi}\int_\G\rmd z\,\frac{\cC_{\textsc{hd}}'(z)}{\cC_{\textsc{hd}}(z)}\nn
&=&\frac{1}{2\pi\rmi}\int_\G\rmd z\,\frac{\cA(-z)+\cA'(-\lst^2z)z-\om^2\cB'(-z)}{\cA(-z)z-\om^2\cB(-z)}\,,\nn\label{argpri}
\ea
where $\G$ is a contour (closed path) in the $z$-plane encircling the $z=\Om_i^2$ roots with $\Re\,\cA_i<0$ and not passing through any zero of $\cC_{\textsc{hd}}(z)$. If we order the surviving roots to be the last ones in the sequence, then the sum in \Eq{solgenfin} runs up to $i=N_+\coloneqq N-N_-$, setting $c_i^\pm=0$ for $i=N_+,N_++1,\dots,N$. 

According to the diffusion method in the passive view, the initial condition for diffusion is at $r_*=0$ and the initial field configuration $X(r=r_*=0,t)$ for the diffusion equation in the $r$-direction satisfies a local equation of motion. The slice position $r_*$ is then shifted along the $r$-direction to the infinite in the limit $r_*\to+\infty$. On this terminal slice, by construction the field $X$ of the localized dynamics must be equal to the field $x$ of the nonlocal dynamics and the two dynamics must coincide according to \Eq{x2}.
Call
\be
\Om_{0i}^2\coloneqq \lim_{r_*\to 0}\Om_i^2(r_*)\,,\qquad \Om_{* i}^2\coloneqq \lim_{r_*\to +\infty}\Om_i^2(r_*)\,.
\ee
At the initial configuration $r_*=0$, the characteristic polynomial \Eq{masterlainf} of the higher-derivative system reduces to $\cC_{\textsc{hd}}(\Om_{0i}^2)= \cA(-\Om_{0i}^2)\Om_{0i}^2=0$ and its roots are
\be\label{roots0}
\Om_{01}^2=0\,,\qquad \big\{\Om_{0i}^2\,:\,\cA(-\Om_{0i}^2)=0,\, i=2,\dots,N\big\}\,.
\ee
At the final configuration $r_*=+\infty$, the roots of $\cC_{\textsc{hd}}(\Om_{*i}^2)= \cA(-\Om_{*i}^2)\Om_{*i}^2-\om^2\cB(-\Om_{*i}^2)=0$ are the $N+1$ constants $\Om_{* i}^2$. At small but finite $r_*$ [small $\la\coloneqq\om^2 f(r_*)\simeq\om^2 r_*$], the roots $\Om_i^2(r_*)$ can be approximated as
\ba
\hspace{-.8cm}r_*\to 0\,:\qquad&&\Om_i^2(r_*)=\Om_{0i}^2+\a_1\,\om^2 r_*+O(\om^4r_*^2)\,,\nn
\hspace{-.8cm}&& \a_1=\left.\frac{\cB(-z)}{\cA(-z)+z\cA'(-z)}\right|_{z=\Om_{0i}^2},\label{asi}
\ea
where $'=\p_z$ and we evaluated $\rmd\cC_{\textsc{hd}}/\rmd\la=\cC_{\textsc{hd}}'\p_{\la}z+\p_{\la} \cC_{\textsc{hd}}=0$ at $\la=0$ and solved for $\a_1=\p_{\la}z|_{\la=0}$. In particular,
\ba
\hspace{-1.cm}r_*\to 0\,:\qquad&&\Om_1^2(r_*)\simeq\frac{\cB(0)}{\cA(0)}\,\om^2 r_*\,,\nn
\hspace{-1.cm}&& \Om_{i\neq 1}^2(r_*)\simeq\left.\frac{\cB(-z)}{z\cA'(-z)}\right|_{z=\Om_{0i}^2}\om^2 r_*\,,\label{rootsasi}
\ea
where $\cA(0)\neq 0$ by hypothesis of nondegeneracy of the roots.


\subsection{Nonlocal dynamics}

In infrared nonlocal gravity, the introduction of an auxiliary field makes the dynamics local similarly to what we did in Sec.~\ref{locdy}, compensating the $\B^{-n}$ terms with $\B^n$ operators. For example, instead of the Lagrangian $\cL\propto R\,[1+f(\B^{-1}R)]$ one can consider
$\tilde\cL\propto R\,[1+f(\phi)]+\psi(\B\phi-R)$. Inverting the equation of motion $\de\tilde\cL/\de\psi=\B\phi-R=0$ with respect to $\phi$ and plugging this back into $\tilde\cL$ yields $\cL$ \cite{Nojiri:2007uq}, were it not for the fact that the solution of this equation is actually $\phi=\B^{-1}R+\la$, where $\la$ is a scalar obeying the homogeneous equation $\B\la=0$ \cite{Koshelev:2008ie}. The extra mode $\la$ enlarges the space of solutions and the parent system $\tilde\cL$ includes, but is not equivalent to, the nonlocal one $\cL$. Exactly the same happens for the higher-derivative system in Sec.~\ref{locdy} and we now have to project out all the solutions artificially introduced above.

Since $\cA(-\Om_i^2)\neq 0$ for $r_*\neq 0$, it is clear that the roots of \Eq{masterlainf} are also roots of the characteristic polynomial obtained by plugging \Eq{solgenfin} into the nonlocal equation \Eq{eom2}:
\be\label{masterlainf2}
\cC_{\textsc{nl}}(\Om_i^2)\coloneqq \frac{\cC_{\textsc{hd}}(\Om_i^2)}{\cA(-\Om_i^2)}=\Om_i^2-\om^2f\frac{\cB(-\Om_i^2)}{\cA(-\Om_i^2)}=0\qquad \forall\,i\,.
\ee
At the initial configuration $r_*=0$, the only root is $\Om_{01}^2=0$, while the other $N-1$ roots in \Eq{roots0} are not present. As soon as $r_*>0$, \Eqq{masterlainf2} acquires $N$ roots, one near each zero of $\cA$ plus one near 0. The asymptotic expansions are as in \Eq{rootsasi}.

The crucial point is that the roots $\Om_i^2(r_*)$ are analytic, smooth functions of $r_*$ and that the latter increases continuously from 0 to $+\infty$. Therefore, in the nonlocal system characterized by \Eq{masterlainf2} the only valid starting point is the root $\Om_{01}^2=0$, which means that the only root of \Eq{masterlainf2} for all $r_*\geq 0$ is $\Om_1(r_*)$. All the other roots $\Om_i(r_*)$ with $i=2,\dots,N$ are artifacts of the higher-derivative system which do not solve \Eq{masterlainf2} at the starting point $r_*=0$, due to \Eq{roots0}. 

It is easy to show that the root $\Om_1(r_*)$ is the one with highest $\Re\,\cA_i>0$. Indeed, from \Eq{asi} we have $z_i=z_{0i}+\a_1\la+O(\la^2)$ and
\ba
\cA_i&=&\cA(-z_i)\nn
&=&\cA(-z_{0i})+ \cA'(-z_{0i})\a_1\la+O(\la^2)\,.\qquad
\ea
For the $i=1$ root, $\Re\,\cA_1=\cA_1\simeq \cA(0)+ \cA'(0)\a_1\la$ grows linearly in $\la$ as $r_*$ increases. For the other roots near $z_{0i}$ where $\cA(-z_{0i})=0$, $\cA(-z_i)\simeq\cB(-z_{0i})\la/z_{0i}\ll 1$ remains small until 
$\la$ becomes large. Therefore, the small-root branch $z_1(\la)$ has $\Re\,z_1(\la)>0$ earliest and grows monotonically with $\la$. Initially, it has strictly larger $\Re\,\cA_i$ than any other branch. Continuous dependence of $\Re\,\cA_i$ from $\la$ implies that, as $\la$ varies over the interval $[0,\om^2]$, the $\Re\,\cA_i$ order of the branches can only change if two branches attain the same real part at some intermediate $\la\in(0,\om^2)$, which can happen only if two roots coalesce (a multiple root; this is excluded by the simplicity assumption, which can be relaxed anyway without changing the main conclusion). Thus, for all $\la$ (all $r_*$) there is a natural ordering 
\be
\Re\,\cA_1\geq\Re\,\cA_2\geq\dots\geq \Re\,\cA_N
\ee
such that $\Om_1^2(r_*)$ is the root with the largest $\Re\,\cA_i>0$ and lies on the analytic continuation of the small branch $\tilde\Om_1^2(r_*)=\om^2 r_*\cB(0)/\cA(0)$ in \Eq{rootsasi}.

Heuristically, at $r_*=0$ only the $z=0$ root exists. As $r_*$ increases, the coupling term $\la\cB/\cA$ turns on in $\cC_{\textsc{nl}}(z)=z-\la\cB/\cA=0$, pushing this root away from 0 while inducing other roots. The root $z_1$ that emerges from 0 naturally evolves towards the one for which the damping rate $\Re\,\cA_i$ is largest, because that branch experiences the strongest balance between the linear term and the $\cB/\cA$ term in $\cC_{\textsc{nl}}$. In dynamical terms, the root $z_1$ with largest damping rate is the least stable, the dominant mode, i.e., the one excited as soon as $r_*$ departs from 0.

The main argument expounded below \Eqq{masterlainf2} leads us to the attractively simple conclusion that an intuitive criterion to select which of the $N$ roots $\Om_{*i}^2$ is the only root of the nonlocal system is perturbativity in $\om$. In the perturbative limit $\om\to 0$, only one root tends to zero [since the higher-order coefficient in the higher-derivative equation is independent of $\om$; see \Eq{roots0}], while the others tend to some finite value $\Om_{0i}^2$:
\bs\label{percri}\ba
&&\Om_1^2(\om^2)\stackrel{\om\to 0}{\longrightarrow} 0\,,\,\qquad \cA_1\longrightarrow \cA(0)>0\,,\\
&&\Om_i^2(\om^2)\stackrel{\om\to 0}{\longrightarrow} \Om_{0i}^2\,,\qquad\!\!\!\! \cA_i\longrightarrow \cA(-\Om_{0i}^2)=0\,,\nn
&&i=2,\dots,N\,.
\ea\es
While in general the system can admit non-perturbative solutions with a finite $\om\neq 0$, the Balakrishnan--Komatsu representation \Eq{posiF} requires that $\Re\,\cA>0$ in \emph{all} its spectrum, independently of the value of $\om$. In particular, from this it follows that not only must the term $\om^2 F$ in the equation of motion tend to a well-defined limit when $\om\to 0$ (which it actually does for all roots $\Om_i^2$ of the higher-derivative parent system), but also $F$ separately, which can only happen if the representation of $F$ is well-defined, i.e., only on the root $\Om_1^2$. 

This only leaves $\Om_1^2$ as the sole root of the characteristic equation, so that the solution of the localized system is the $i=1$ mode of \Eq{solgenfin}. If we further specify the $r_*$-dependence of the coefficients $c_1^\pm(r_*)=\exp(r_*\cA_1) \tilde c_1^\pm(r_*)$, where $\tilde c_1^\pm(r_*)$ are such that $\tilde c_1^\pm(+\infty)$ are finite, then we have
\be
X(r,t)=\rme^{(r_*-r)\cA_1}\left[\tilde c_1^+(r_*)\,\rme^{\rmi\Om_1(r_*) t}+\tilde c_1^-(r_*)\,\rme^{-\rmi\Om_1(r_*) t}\right].
\ee
This profile is ill-suited to obtain the nonlocal solution $x(t)$ in the active view \Eq{x1}, since for any fixed $r$ its normalization blows up when we remove the cutoff $r_*\to+\infty$. However, it is exactly what we need in the passive view \Eq{x2}. When $r=r_*$, the normalization becomes 1. When $r_*\to 0$, we get 
\ba
X(0,t)&=&\lim_{r_*\to 0} X(r_*,t)\nn
&=&[\tilde c_1^+(0)+\tilde c_1^-(0)]\nn
&&+\rmi [\tilde c_1^+(0)-\tilde c_1^-(0)]\Om_{01}^2t+O(\Om_{01}^2)\,,\qquad
\ea
and we recover the initial configuration \Eq{inicon}
provided $\tilde c_1^\pm(0)=(a_0\Om_{01}\mp \rmi a_1)/(2\Om_{01})$. This expression is immediately promoted for all $r_*$ to
\bs\ba
&&\tilde c_1^\pm(r_*)
=\frac{a_0}{2}\mp \frac{\rmi a_1}{2\Om_1(r_*)}\,,\\
&& c_{*1}^\pm\coloneqq \tilde c_1^\pm(+\infty)
=\frac{a_0}{2}\mp \frac{\rmi a_1}{2\Om_{*1}}\,,
\ea\es
which is well-defined when $r_*\to+\infty$. The solution of the nonlocal dynamics is thus
\ba
x(t)&=&\lim_{r_*\to+\infty} X(r_*,t)\nn
&=&c_{*1}^+\,\rme^{\rmi\Om_{*1} t}+c_{*1}^-\,\rme^{-\rmi\Om_{*1} t}\nn
&=&a_0\cos(\Om_{*1} t)+\frac{a_1}{\Om_{*1}}\sin(\Om_{*1} t)\,.\label{physol}
\ea


\subsection{Examples}

Let us present a few examples some of which were considered also in \cite{Anselmi:2025uda}. The aim is not to test the diffusion method directly but to use the perturbativity criterion \Eq{percri} as a quick way to select the physical frequency $\Om_{*1}^2$ of the solution \Eq{physol}.

\subsubsection{Tachyonic fakeon two-derivative models}

In this double example \cite[Secs.~4.1.2 and 4.1.3]{Anselmi:2025uda},
\be
F(\p_t) = \frac{1}{1-\lst^2\p_t^2}\,,\qquad \cA=1-\lst^2\p_t^2\,,\qquad \cB=\pm 1\,,\label{posiF2}
\ee
where $\lst$ is a length-time scale ($[\lst]=-1$) denoted as $\tau$ in \cite{Anselmi:2025uda} and the sign of $\cB$ determines whether the force acting on $x(t)$ is attractive or repulsive. The master equation \Eq{masterlainf} becomes
\ben
\cC_{\textsc{hd}}(\Om_i^2)=\lst^2\Om_i^4+\Om_i^2\mp\om^2=0\,,
\een
which has $N=2$ roots $\Om_i^2$:
\bs\label{roo2}\ba
&&\Om_1^2(r_*)=\frac{\sqrt{1\pm4\lst^2\om^2f(r_*)}-1}{2\lst^2}\,,\\
&& \Om_2^2(r_*)=\frac{-\sqrt{1\pm4\lst^2\om^2f(r_*)}-1}{2\lst^2}\,.\qquad
\ea\es
In the model with attractive force ($+$ sign), only $\Om_1^2$ obeys \Eq{posiA}, $\Re\,\cA_1> 0$, in agreement with the result of \cite{Anselmi:2025uda}. This is sufficient by itself to remove the unphysical root $\Om_2^2$. In the model with repulsive force [$-$ sign in \Eq{roo2}], both roots have $\Re\,\cA_i> 0$ and \Eq{posiA} is not enough to superselect the physical root. Here is where the perturbativity criterion \Eq{percri} comes into play. Indeed, in the limit $\om f\to 0$ one has $\Om_1^2\to 0$ and $\Om_2^2\to -1/\lst^2$ in both cases, signaling that $\Om_2^2$ approaches the unphysical root of $\cA(1/\lst^2)=0$. 

\subsubsection{Standard fakeon two-derivative model}

The toy model of fakeons with real masses \cite[Sec.~4.1.4]{Anselmi:2025uda} requires a specific representation for the operator $F$, which we write as the regularization
\ba
&&F(\p_t) = \frac{1}{1+\lst^2\p_t^2}\coloneqq \lim_{\ve\to 0}\frac{1+\lst^2\p_t^2}{\ve^2+(1+\lst^2\p_t^2)^2}\,,\nn
&& \cA=\ve^2+(1+\lst^2\p_t^2)^2,\qquad \cB=1+\lst^2\p_t^2\,,\qquad\qquad
\ea
where the limit $\ve\to 0$ does not commute with the integration in $\t$ and must be taken at the very end. 
The master equation \Eq{masterlainf} for finite $\ve$ becomes
\ben
\cC_{\textsc{hd}}(\Om_i^2)=[(\lst^2\Om_i^2-1)^2+\ve^2]\Om_i^2+\om^2(\lst^2\Om_i^2-1)=0\,,
\een
which has $N=3$ roots $\Om_i^2$: a real one and two complex-conjugate ones. The condition \Eq{posiA} is trivially satisfied by the real root $\Om_1^2$ and discards the two complex roots with a nonvanishing imaginary part, for which $\Re\,\cA<0$. This is in agreement with the criterion \Eq{percri}, since $\Om_1^2\to 0$ in the perturbative limit $\om\to 0$ while the other roots $\Om_2^2$ and $\Om_3^2$ tend to the roots $(1\pm\rmi)/\lst^2$ of $\cA=0$.

For instance, for $\lst=1=\om$ and $\ve=1$ we have three solutions $\Om_1\approx 0.66$, $\Om_{2,3}\approx 1.07\pm 0.61\rmi$, in agreement with the direct calculation of \cite{Anselmi:2025uda}. In the limit $\ve\to 0$, $\Om_1=1$ and $\Om_{2,3}\approx 0.87\pm 0.5\rmi$; the complex roots are thrown away.

\subsubsection{Higher-order fakeon model}

Another example has a tenth-order parent model with
\be
\cA=1+(\lst^2\p_t^2)^4\,,\qquad \cB=1+(\lst^2\p_t^2)^2\,.
\ee
One root $\Om_1^2$ of the characteristic equation of the higher-derivative system is real, while the other four are organized into two complex-conjugate pairs. Depending on the value of $\lst$ and $\om$, one or even both pairs can violate \Eq{posiA}. In the limit $r_*\to 0$, the real root tends to zero and is actually the one in the solution of the nonlocal system; the paired roots collapse to the roots $\Om_{0i}^2=\rme^{\rmi\pi(2i+1)/4}/\lst^2$ of $\cA_i=0$, $i=2,3,4,5$, and are therefore excluded.


\section{Discussion}\label{sec5}

In this paper, we have studied the Cauchy problem of scalar-field systems with fakeonic (i.e., purely virtual) modes and found that one needs two initial conditions to fully specify the nonlocal classicized dynamics after fakeons are projected out. This gives not only an independent confirmation of the general results and of the examples of \cite{Anselmi:2025uda} but also a systematic way to find solutions of both nonlinear and linear systems. We followed step by step the diffusion method and applied it to fakeonic systems for the first time. For linear systems, we gave the rationale for why the only two independent solutions of the nonlocal equation of motion are those whose frequencies $\pm\Om_{*1}$ tend to zero in the perturbative limit of vanishing coupling $\om\to0$. This identification of the solutions of the nonlocal system with the perturbative ones was observed also in \cite{Anselmi:2025uda}. From the perspective of the diffusion method, it is a consequence of the fact that, on one hand, the starting point of diffusion is a slice where the dynamics happens to be the perturbative one and, on the other hand, this slice is smoothly connected with the end point of diffusion where one recovers the original nonlocal dynamics.

\begin{figure}[ht]
	\bc
	\includegraphics[width=8.5cm]{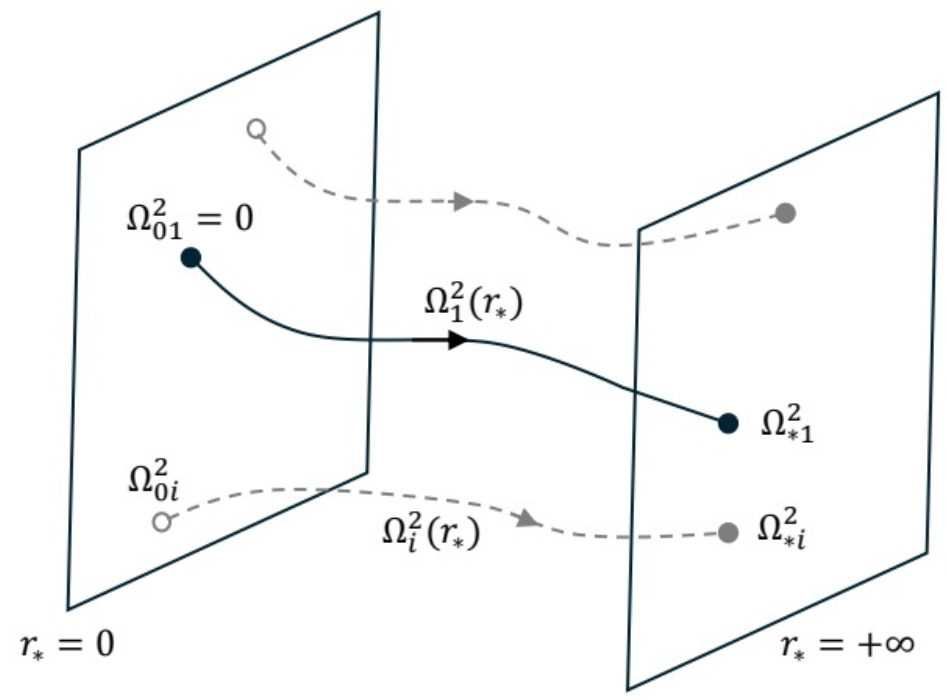}
	\ec
	\caption{\label{fig1} On the slice $r_*=0$, only the root $\Om_{01}^2=0$ (filled dot) is a solution of the characteristic equation \Eq{masterlainf2} of the linear nonlocal system, while the roots of $\cA(\Om_{0i}^2)=0$ are not (empty dots). On the slice $r_*=+\infty$, all the roots $\Om_{*i}^2$ of the higher-derivative parent system are solutions of \Eq{masterlainf2} (filled dots) but only $\Om_{*1}^2$ is smoothly connected to $\Om_{01}^2=0$ by the function $\Om_1^2(r_*)$ (solid curve). Since the direction of diffusion is only towards increasing values of the extra direction, it is not possible to ``go upstream'' from any slice with finite $r_*>0$ back to the $r_*=0$ slice. This selects $\Om_{*1}^2$ as the only physical root of the nonlocal system.}
\end{figure}

All of this was achieved without picking a generalized integral representation of the nonlocal operator as done in \cite{Anselmi:2025uda}. Here we only used the coordinate-independent Balakrishnan--Komatsu representation \Eq{bkrep} of the inverse of an operator and found that it correctly encodes the fakeon prescription. The price to pay for this simplification is that it requires a careful formulation of the diffusion process used to understand and solve the nonlocal dynamics. Perhaps the most difficult point to digest in the case of linear systems is that the diffusion equation establishes a one-way direction for the evolution of the localized system, from vanishing $r$ (or $r_*$) to arbitrarily large values. Going backwards in $r$ is simply not possible in this setting, as one can check by attempting to evolve an $r=+\infty$ profile back to $r=0$. This implies that the starting point of diffusion is also the place where one selects the physical frequencies of the solution. Any other frequency, legitimate for the higher-derivative parent system, formally becomes a solution of the nonlocal system as soon as $r_*>0$ but it is an artifact of insisting to run the diffusion flow backwards from some finite $r_*>0$ (or $r_*=+\infty$) to zero. This is depicted in Fig.~\ref{fig1}.

A notable difference of fakeonic models with respect to nonlocal QFT and quantum gravity with entire form factors \cite{Calcagni:2018lyd,Calcagni:2018gke} and to fractional QFT and quantum gravity \cite{Calcagni:2025wnn} is that the inverse of the nonlocal operator appearing in the dynamics is a local operator, hence one can get a higher-derivative parent system with more solutions than the nonlocal one. In turn, this gives more tools to solve the nonlocal dynamics compared with the other nonlocal theories considered in the literature but it also requires adaptations of the diffusion method. In the case of fakeonic systems, nonlocality is fully confined in a term that actually vanishes in the limit of zero nonlocal length-time scale $\lst$, just like in fractional QFT but contrary to nonlocal QFT with entire form factors. This results in the remarkable coincidence, in linear systems, that the initial diffusion slice at $r_*=0$ hosts the perturbative dynamics and solutions in the limit $\om\to 0$. The fact that the surviving root $\Om_1^2$ is the perturbative one does not stem from imposing to solve the system perturbatively (which would be an added assumption) but from the noninvariance under $r$-reversal of the diffusing dynamics, as illustrated by Fig.~\ref{fig1}. This might be perceived as an extra requirement not present in \cite{Anselmi:2025uda} but a careful scrutiny shows that this is not the case. As said above, both here and in \cite{Anselmi:2025uda} a representation of the nonlocal form factor $F$ is chosen under independent criteria: in \cite{Anselmi:2025uda}, to implement the Anselmi--Piva prescription; here, to implement the diffusion equation. The result is the same and no extra requirements are needed in either method. In particular, one-way directionality along $r$ (or $r_*$) is an in-built feature of the diffusion equation, hence of the representation chosen for $F$. Different representations of $F$ can amount to different operators and a different counting of the initial conditions. The diffusion method assumes a certain representation of any given $F$ and, from that and from the one-way directionality of the diffusion flow, it reaches certain conclusions about the Cauchy problem. Asking whether the method misses some solutions because it imposes this one-way directionality is tantamount to demand it to say anything about solutions with different representations of $F$ for which the method itself does not apply. Which is a circular logic.

The diffusion method proposed here is very similar to the one employed in fractional QFT \cite{Calcagni:2025wnn} and lessons learned in one context could have applications to the other, or to other settings with a similar type of nonlocality such as infrared nonlocal gravity \cite{Barvinsky:2003kg,Deser:2007jk,Belgacem:2017cqo,Belgacem:2020pdz}. In particular, here we found that the localization of linear fakeonic systems in a $(D+1)$-dimensional spacetime with a fictitious extra coordinate $r$ works better with a ``passive'' slicing of the extra direction, where one sits in a fixed $r$-slice and the nonlocal dynamics is recovered when this slice is sent to infinity. It will be interesting to explore whether this passive slicing in the diffusion method also has applications in other nonlocal theories.


\begin{acknowledgments}
G.C.\ thanks D.~Anselmi for his perceptive and tireless criticism in many invaluable discussions. The author is supported by Grant No.\ PID2023-149018NB-C41 funded by the Spanish Ministry of Science, Innovation and Universities
MCIN/AEI/10.13039/501100011033.
\end{acknowledgments}

\end{document}